\documentclass[twocolumn,showpacs,preprintnumbers,amsmath,amssymb,unsortedaddress,floatfix,aps,pra]{revtex4}

\usepackage{amsfonts}
\usepackage{mathrsfs}
\usepackage{amsmath}
\usepackage{graphicx}

\newcommand{\ket}[1]{\mbox{$|#1\rangle$}}

\begin{document}

\title{Simple scheme for expanding a polarization-entangled $W$ state by adding one photon}
\author{Yan-Xiao Gong}
\email{yxgong@mail.ustc.edu.cn}
\author{Xu-Bo Zou}
\author{Yun-Feng Huang}
\email{hyf@ustc.edu.cn}
\author{Guang-Can Guo}
\affiliation{Key Laboratory of Quantum Information, University of
Science and Technology of China, CAS, Hefei, 230026, People's
Republic of China}

\date{\today }

\begin{abstract}
We propose a simple scheme for expanding a polarization-entangled
$W$ state. By mixing a single photon and one of the photons in an
$n$-photon $W$ state at a polarization-dependent beam splitter
(PDBS), we can obtain an ($n+1$)-photon $W$ state after
post-selection. Our scheme also opens the door for generating
$n$-photon $W$ states using single photons and linear optics.
\end{abstract}


\maketitle

Entanglement not only plays a central role in fundamental quantum
physics \cite{EPR, Bell_nonlocal}, but also has wide applications in
quantum information processing, such as quantum teleportation
\cite{bennett_tele}, dense coding \cite{bennett_dense}, quantum
cryptography \cite{Ekert_QKD}, and quantum computation
\cite{nielsen_book}. While bipartite entanglement has been well
understood, multipartite entanglement offers a very complicated
structure. For example, it was shown that genuine three-particle
entanglement can be classified into two classes by the equivalence
under stochastic local operations and classical communication
(SLOCC) \cite{Dur_two}. One is the Greenberger-Horne-Zeilinger (GHZ)
state \cite{GHZ_book}
\begin{equation}
\ket{GHZ_3}=\frac{1}{\sqrt{2}}\left(\ket{000}+\ket{111}\right).
\end{equation}
The other is the $W$ state
\begin{equation}
\ket{W_3}=\frac{1}{\sqrt{3}}\left(\ket{001}+\ket{010}+\ket{100}\right).
\end{equation}
The GHZ state is usually taken as ``maximally entangled'' state in
some senses, for instance, it violates Bell inequalities maximally.
However, it is also maximally fragile, i.e., if one or more
particles are lost or discarded, then all the entanglement is
destroyed. The $W$ state is less entangled in the sense that its
violation is weaker than that of GHZ state. While, the $W$ state is
very robust against the loss of one of the particles, namely,
two-particle entanglement can be observed after one particle is lost
or measured. Thereby, in this sense, the $W$ state is more
entangled.

The entanglement persistency  property can be easily obtained from
the representation of the $n$-particle $W$ state
\begin{align}\label{wdefine}
\ket{W_n}=&\frac{1}{\sqrt{n}}(\ket{00\cdot\cdot\cdot01}+\ket{00\cdot\cdot\cdot10}+\cdot\cdot\cdot\nonumber\\
&+\ket{01\cdot\cdot\cdot00}+\ket{10\cdot\cdot\cdot00})\nonumber\\
=&\frac{1}{\sqrt{n}}\ket{n-1,1},
\end{align}
where $\ket{n-1,1}$ denotes the (unnormalized) totally symmetric
state including $n-1$ particles in state $\ket{0}$ and one particle
in state $\ket{1}$,
e.g.,$\ket{3,1}=\ket{0001}+\ket{0010}+\ket{0100}+\ket{1000}$. We can
see that any particle is entangled with the other particles and that
all the particles are equivalent. In fact, it was shown that the $W$
state has the maximum degree of entanglement between any pair of
particles \cite{W_maximal}. These interesting features lead the
$W$-class states to applications in a variety quantum information
processing tasks, such as quantum teleportation
\cite{W_teleportation1, W_teleportation2, W_teleportation3}, dense
coding \cite{W_dense}, quantum secret communication \cite{W_QKD}.

Linear optical systems have supplied a broad field for experimental
implementation of multipartite entangled states. There have  been
many proposals \cite{W_gen0210, W_gen0212, W_gen0304, W_gen0407,
W_gen0411, W_gen0503, W_gen0702} and experimental implementations
\cite{W_exp0304, W_exp0402, W_exp0510, W_exp0802} for producing $W$
states. Quite recently, Tashima \emph{et al.} introduced an
interesting optical gate for expanding polarization-entangled $W$
states \cite{W_expand}. In their scheme, after the operation of the
gate on one of the photons in an $n$-photon $W$ state, an
\mbox{$(n+2)$}-photon $W$ state can be obtained after
post-selection.

In this paper, using a similar expanding principle with that in Ref.
\cite{W_expand}, we propose a simple scheme for expanding a
polarization-entangled $W$ state by adding a single photon to the
existing state, rather than adding two photons in Ref.
\cite{W_expand}. Our scheme needs only a polarization-dependent beam
splitter (PDBS), where one of the photons in an $n$-photon $W$ state
interferences with a single photon and after post-selection an
\mbox{$(n+1)$}-photon $W$ state can be obtained.

\begin{figure}[tb]
\centering
\includegraphics[width=4cm]{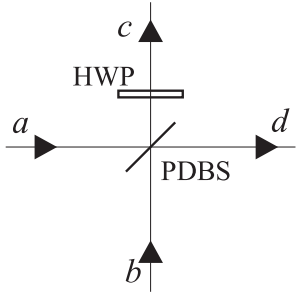}
\caption{Scheme for adding a single photon to an $n$-photon $W$
state. The polarization-dependent beam splitter (PDBS) has
reflectivities of \mbox{$\eta_H=(5-\sqrt{5})/10$} and
\mbox{$\eta_V=(5+\sqrt{5})/10$} for horizontally ($H$) and
vertically ($V$) polarized photons, respectively. One of the photons
in the $W$ state is input in mode $a$, and a single photon in state
$\ket{H}$ is added in mode $c$. A half-wave plate (HWP) oriented at
$0^{\circ}$ can introduce a phase shift of $\pi$ between $H$ and $V$
polarized photons. This scheme succeeds in the case of twofold
coincidence detection in the output modes $c$ and $d$.} \label{wfig}
\end{figure}

Before introducing our scheme we would like to note that the qubits
here are all encoded in polarization states of single photons, so
that $\ket{0}\equiv\ket{H}$ and $\ket{1}\equiv\ket{V}$, where
$\ket{H}$ ($\ket{V}$) denotes the horizontal (vertical) polarization
state. Our scheme for adding a single photon to an $n$-photon $W$
state is depicted in Fig.~\ref{wfig}. The key of our scheme is an
element of PDBS, with reflectivities of
\begin{equation}
\eta_H=\frac{5-\sqrt{5}}{10}\ \ \ \ \ \ \text{ and }\ \ \ \ \ \
\eta_V=\frac{5+\sqrt{5}}{10},
\end{equation}
for horizontally ($H$) and vertically ($V$) polarized photons,
respectively. Such class of elements has been used in several
experiments \cite{PDBS_cluster, PDBS_CNOT1, PDBS_CNOT2, PDBS_CNOT3}.
One of the photons in state $\ket{W_n}$ and a single photon in state
$\ket{H}$ meet at the PDBS, which are input in modes $a$ and $b$,
respectively. If they are indistinguishable except the degrees of
path and polarization (fourth-order interference will happen for the
same polarization photons), the state transformations at the PDBS
can be expressed as
\begin{align}
\ket{H}_a\rightarrow&\sqrt{\eta_H}\ket{H}_c+\sqrt{1-\eta_H}\ket{H}_d,\\
\ket{V}_a\rightarrow&\sqrt{\eta_V}\ket{V}_c+\sqrt{1-\eta_V}\ket{V}_d,\\
\ket{H}_b\rightarrow&\sqrt{1-\eta_H}\ket{H}_c-\sqrt{\eta_H}\ket{H}_d.
\end{align}
After the PDBS, we use a half-wave plate (HWP) set to $0^{\circ}$ to
introduce a phase shift of $\pi$ between $H$ and $V$ polarized
photons, with the transformations
\begin{equation}
\ket{H}_c\rightarrow\ket{H}_c,\ \ \ \ \ \ \ \ \
\ket{V}_c\rightarrow-\ket{V}_c.
\end{equation}
Therefore, if we post-select the successful events, i.e., twofold
coincidence detection at the output modes $c$ and $d$, we can obtain
the state transformations as follows,
\begin{align}\label{expand_1}
\ket{H}_a\ket{H}_b\rightarrow&\frac{1}{\sqrt{5}}\ket{H}_c\ket{H}_d,\\
\label{expand_2}\ket{V}_a\ket{H}_b\rightarrow&\frac{1}{\sqrt{5}}\left(\ket{H}_c\ket{V}_d+\ket{V}_c\ket{H}_d\right).
\end{align}

Next we explain how a single photon can be added to a state
$\ket{W_n}$ through our scheme. Since all the photons in the $W$
state are equivalent, we can choose any photon to inject in mode
$a$, for instance, mode $n$, so that we can rewrite the $W$ state
given by Eq.~(\ref{wdefine}) as follows,
\begin{align}
\ket{W_n}=&\frac{1}{\sqrt{n}}\ket{n-1,1}\nonumber\\
=&\frac{1}{\sqrt{n}}\left[\ket{n-2,1}\ket{H}_n+\ket{H}^{\otimes^{(n-1)}}\ket{V}_n\right]\nonumber\\
\longrightarrow&\frac{1}{\sqrt{n}}\left[\ket{n-2,1}\ket{H}_{a}+\ket{H}^{\otimes^{(n-1)}}\ket{V}_a\right].
\end{align}
Then we can write the state evolution of the photon-added process as
\begin{align}\label{wend}
\ket{W_n}\ket{H}_b\rightarrow&\frac{1}{\sqrt{5n}}\Big{[}\ket{n-2,1}\ket{H}_c\ket{H}_d+\ket{H}^{\otimes^{(n-1)}}\nonumber\\
&\otimes\big{(}\ket{H}_c\ket{V}_d+\ket{V}_c\ket{H}_d\big{)}\Big{]}+\ket{\Phi}\nonumber\\
=&\sqrt{\frac{n+1}{5n}}\ket{W_{n+1}}+\ket{\Phi},
\end{align}
where $\ket{\Phi}$ is an unnormalized state including the amplitudes
that would not lead to the successful events.

\begin{figure}[tb]
\centering
\includegraphics[width=7cm]{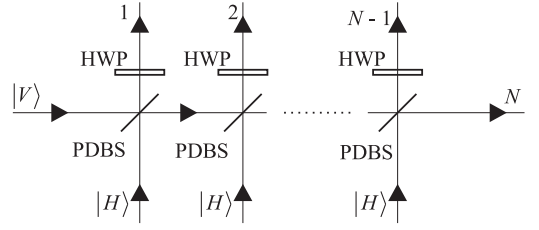}
\caption{Schematic of preparing an $N$-photon $W$ state using single
photons and the scheme shown in Fig.~\ref{wfig}.} \label{wfig2}
\end{figure}

From Eq.~(\ref{wend}) we can see that the success probability for
adding a single photon to a state $\ket{W_n}$ is
\mbox{$(n+1)/(5n)$}, which approaches a constant $1/5$ when $n$
becomes large. It is not difficult to find that if we generate an
$N$-photon $W$ state using single photons and our scheme (see
Fig.~\ref{wfig2} for the schematic), the total success probability
is \mbox{$N/5^{N-1}$ $(N\geqslant2)$}. However, this is not optimal.
By replacing the leftmost PDBS with a balanced
polarization-independent beam splitter, namely,
\mbox{$\eta_H=\eta_V=1/2$}, the probability of success can be
improved to \mbox{$N/(4\times5^{N-2})$}. Alternatively, if we do not
restrict our sources to single photons and EPR states are available,
we can get higher probability. Explicitly, with our scheme, we can
first add a single photon to an EPR state
\mbox{$(\ket{HV}+\ket{VH})/\sqrt{2}$} to get state $\ket{W_3}$ and
then add single photons one by one as the way in Fig.~\ref{wfig2}.
In this case we can obtain the state $\ket{W_N}$ with the success
probability \mbox{$N/(2\times5^{N-2})$}. In particular, the success
probability for preparing state $\ket{W_3}$ is $3/10$, which is
highest compared with other linear optical schemes, as the most
efficient one at present is $3/16$ in Ref. \cite{W_expand}. The
success probability for preparing state $\ket{W_4}$ is $2/25$, which
is lower than $1/8$ in Ref. \cite{W_expand} but still higher than
the other linear optical schemes (the most efficient one before the
scheme in Ref. \cite{W_expand} is $2/27$ in Ref. \cite{W_gen0210}).
Therefore, we believe our scheme is experimental feasible for
preparing states $\ket{W_3}$ and $\ket{W_4}$. However, as the number
of photons increases the success probability decreases
exponentially, so experimental preparing more-photon $W$ states
would be still difficult. Actually, this is a common problem in many
linear optical schemes.

Finally, we would like to give a brief discussion on the comparison
of our scheme with the scheme in Ref. \cite{W_expand}. The two
schemes are based on similar expanding principles (this can be seen
from Eqs. (\ref{expand_1}) and (\ref{expand_2}) and Eqs. (3) and (4)
in their scheme), but a single photon is added in our scheme rather
than two photons are added in their scheme. This leads to different
experimental requirements, i.e., we need single photons while they
need two-photon Fock states. In their scheme, the success
probability is \mbox{$(2k+1)2^{-4k}$} for preparing state
$\ket{W_{2k+1}}$ and \mbox{$(k+1)2^{-4k}$} for preparing state
$\ket{W_{2(k+1)}}$. Therefore, our scheme is more efficient for
preparing state $\ket{W_3}$ but less efficient for preparing
more-photon $W$ state.

In conclusion, we propose a simple scheme to expand a
polarization-entangled $W$ state by adding a single photon. This
method should be very helpful in quantum information processing in
the future when quantum memory and nondemolition measurements are
available, because we can add a single photon to an existing $W$
state easily to get a larger one. Furthermore, our method gives a
new way to prepare $W$ state using single photons and linear optical
elements.

This work was funded by National Fundamental Research Program (Grant
No. 2006CB921907), National Natural Science Foundation of China
(Grants No. 60621064, No. 10674128 and No. 10774139), Innovation
Funds from Chinese Academy of Sciences,`` Hundreds of Talents''
program of Chinese Academy of Sciences, Program for New Century
Excellent Talents in University, A Foundation for the Author of
National Excellent Doctoral Dissertation of PR China (grant 200729).

\end{document}